\documentclass{article}
\usepackage{epsfig}
\begin{document}
\newcommand{\ccdot}{ \! \cdot \! }
\hfill  INFNNA-IV-2001/25
\vspace{1.0cm}
\renewcommand{\thefootnote}{\fnsymbol{footnote}}
\begin{center}
{\LARGE \bf  Radiative rare kaon decays}\footnote[2]{To appear in
 the Proceedings of KAON 2001, Pisa, Italy, June 12-17, 2001}
\end{center}
\vspace{0.8cm}
\centerline{ Giancarlo D'Ambrosio\footnote[1]{On leave of absence at
{\it Theory Division, CERN, CH-1211, Geneva 23 Switzerland}}}
\vspace{0.5cm}
\centerline{\it INFN-Sezione di Napoli, I-80126 Napoli, Italy}
\centerline{\emph{E-mail: Giancarlo.D'Ambrosio@cern.ch}}
\vspace{1cm}


\baselineskip=11.6pt
\begin{abstract}
  We discuss  theoretical issues in  radiative rare kaon decays.
The interest is twofold: to extract useful short-distance information
and understand the underlying dynamics. 
We emphasize channels where either we can understand non-perturbative 
aspects of QCD or  there is a chance to test the Standard Model.
\end{abstract}
\baselineskip=14pt
\section{Introduction}

Kaon decays are an important place to study non-perturbative aspects of 
QCD and test  the Standard Model.  Indeed
some  channels are completely 
dominated by long-distance dynamics, such as the
 CP-conserving amplitude for
$K \rightarrow \pi \pi$ and 
others, like  
$K\rightarrow \pi \nu \overline{\nu }$ \cite{buchalla,buras01}, 
which are described in
 terms of pure short-distance physics.
In this review we will be mostly concerned with kaon decays involving
electromagnetic interactions and thus long-distance phenomena are 
not negligible. However, as we shall see, it is still  possible in these 
channels
 to
extract the short-distance component with an accurate analysis.
Indeed there are plenty of motivations to look for new physics (NP) in
these kaon decays \cite{NP}. 
\renewcommand{\thefootnote}{\arabic{footnote}}
The channels which will be considered here are
 $K_{S}\rightarrow \gamma \gamma $, $K\rightarrow \pi \gamma \gamma$,
$K\rightarrow \pi \ell ^{+}\ell^{-}$, $K\rightarrow \pi \pi \gamma $ and  
$K_{L}\rightarrow \mu \overline{\mu }$.
Experiments at CERN, Fermilab, Brookhaven, KLOE
 \cite{NA48kpgg}-\cite{NA48KSb} 
are and will be also providing a large amount of
data to  further motivate this research.
 QCD at low energy will be studied in the
 framework of chiral perturbation theory
(ChPT) \cite{Weinberg1}-\cite{Eckerp}. 
The $\Delta S=1$ weak Lagrangian is
expanded in powers of   external momenta and masses:
there is only one ${\cal O}(p^2)$
operator for the $\Delta I=1/2$ and 
$\Delta I=3/2$, with coefficients    $G_8$ and $G_{27}$ respectively, 
determined from  $K\rightarrow \pi \pi$
transitions.
 The  ${\cal O}(p^4)$ Lagrangian has many operators $W_{i}$, 
and corresponding coefficients $N_{i}$ \cite{EKW93}:
\begin{equation}
{\cal L}^{(p^4)}_{\Delta S=1}=G_8 F^2 \sum_{i=1}^{37}N_{i} W_{i}
\label{eq:wp4}
\end{equation}
and although there are already  interesting tests at this level, 
as we shall see, it is clear  that further assumptions are needed  in
 order to 
be reasonably predictive, typically of vector meson dominance and 
1/$N$ \cite{EKW93}-\cite{derafael}.

\section{$K_{S}\rightarrow \gamma \gamma $}
$K_S\rightarrow\gamma\gamma$ has vanishing short-distance 
contributions \cite{GL}, so it is 
a pure 
long-distance phenomenon; since the external particles are neutral there is no 
${\cal O }(p^2)$ amplitude. For the same reason, if we write down the
 ${\cal O}(p^4)$ counterterm structure,
$F_{\mu\nu}F^{\mu\nu} \langle\lambda_6 Q U^+ Q
U\rangle$, this gives a vanishing contribution 
(we use the standard chiral 
notation as in Ref. \cite{DI98});
\begin{figure}[htb]
\vfill
\begin{minipage}[b]{5cm}\centering
\mbox{\epsfig{file=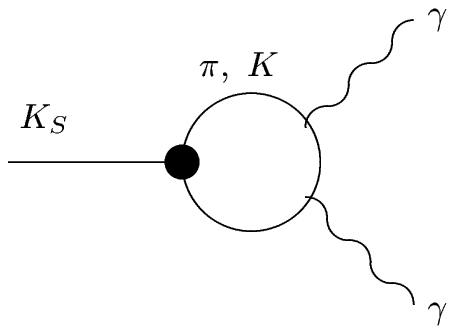,width=4cm,height=3cm}}
\caption{$K_{S}\rightarrow \gamma \gamma $ \cite{DEG}} \label{fig:A}
\end{minipage}
\hspace{1.0cm}
\begin{minipage}[b]{5.5cm}\centering
\mbox{\epsfig{file=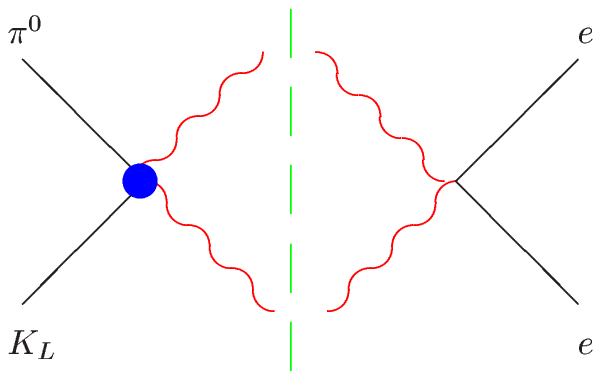,width=5cm,height=3cm}}
\caption{CP-conserving contribution to $K_{L}\rightarrow \pi ^0 e^+ e^- $}
\label{fig:klpee}
\end{minipage}
\vfill
\end{figure}
this implies that at ${\cal O}(p^4)$:  i) we 
have only  a loop contribution in Fig. \ref{fig:A}
 and ii) this contribution is scale-independent;
 it is completely predicted
by the  $K_{S}\rightarrow \pi \pi $ amplitude  \cite{DEG}
and can be compared with the recent
NA48 result \cite{NA48Ksgg}:
\begin{equation}
B(K_S\rightarrow\gamma\gamma)_{\rm Ch PT}=2.1\times 10^{-6}
\quad
B(K_S\rightarrow\gamma\gamma)_{\rm NA48}=(2.6\pm 0.4)\times 10^{-6}.
\label{eq:Ksgg}
\end{equation}
${\cal O }(p^6)_{\rm CT}$, as the structure
\begin{equation}
F^{\mu\nu}F_{\mu\nu}\langle\lambda_6 Q^2 \mu M
U^+\rangle \quad \label{eq:Ksp6},
\end{equation}
in principle  can  modify Eq.(\ref{eq:Ksgg}), but 
 chiral power counting suggests
 $ A^{(6)}/A^{(4)}\sim
m_K^2/(4\pi F_\pi)^2\sim 0.2$. In fact the potentially large 
Vector Meson (VMD) contributions, which could alter this relation as in 
$K_L\rightarrow \pi^0 \gamma\gamma$, are absent in this channel.
Since higher order $\pi$-loop corrections are small \cite{KH94},
we can look at the theoretical prediction in  Eq.(\ref{eq:Ksgg})
as a test of non-VMD contributions.

\section{$K\rightarrow \pi \gamma \gamma$ decays and the
CP-conserving $K_{L} \rightarrow \pi ^0 \ell ^{+}\ell^{-}$ }
$K_{L} \rightarrow \pi ^0 \ell ^{+}\ell^{-}$ is a 
classical example of 
how our
control on low energy theory may help to disentangle 
short-distance physics.
In fact 
the effective current$\otimes $cur\-rent structure of weak interactions
obliges short-distance contributions to $K_{L}\rightarrow \pi ^{0}\ell
^{+}\ell ^{-},$ 
analogously to $K_{L}\rightarrow \pi ^{0}\nu \overline{\nu }
, $ to be direct CP-violating \cite{buchalla,gino98a}. 
However, differently from the
neutrino case, $K_{L}\rightarrow \pi ^{0}\ell ^{+}\ell ^{-}$ 
receives also
non-negligible long-distance contributions:
 i) indirect CP-violating from 
one-photon exchange, discussed in the next section,
 and ii) CP-conserving from two-photon exchange in 
Fig. \ref{fig:klpee}, where the photons can be on-shell (two-photon
 discontinuity) and thus  directely related to the observable 
$K_{L}\rightarrow \pi ^{0}\gamma \gamma$ decay, or off-shell 
and then a form factor should be used \cite{DG95}. 
We will comment in the conclusions on possible ways to avoid the potential
 large
background contribution from 
$K_{L}\rightarrow e^{+}e^{-}\gamma \gamma $ \cite{greenlee}.
 The present bounds
from KTeV \cite{Littenberg,KTeVklpi0ll,diwan} are 
\begin{equation}
B(K_{L}\rightarrow \pi ^{0}e^{+}e^{-})<5.1\times 10^{-10}\quad
{\rm and}\quad
B(K_{L}\rightarrow \pi ^{0}\mu ^{+}\mu ^{-})<3.8\times 10^{-10}.
\end{equation}
\hspace*{0.1cm} The general amplitude for $K_{L}(p)\rightarrow \pi
^{0}\gamma (q_{1})\gamma (q_{2})$ can be written in terms of two 
Lorentz and gauge invariant amplitudes $A(z,y)$ and $B(z,y):$
\begin{eqnarray}
  && {\cal A}( K_{L}\rightarrow \pi ^{0}\gamma \gamma ) =  
   \frac{G_{8} \alpha }{ 4\pi }
   \epsilon_{1 \mu} \epsilon_{2 \nu} 
   \Big[  A(z,y) 
    (q_{2}^{\mu}q_{1}^{\nu }-q_{1}\ccdot q_{2}~ g^{\mu \nu } ) +  \nonumber \\
 && \qquad + 
   \frac{2B(z,y)}{m_{K}^{2}} 
    (p \ccdot q_{1}~ q_{2}^{\mu }p^{\nu} + p\ccdot q_{2} ~p^{\mu }q_{1}^{\nu}
   -p\ccdot q_{1}~ p\ccdot q_{2}~ g^{\mu \nu } 
   - q_{1}\ccdot q_{2}~ p^{\mu }p^{\nu } ) \Big]~, \quad
\label{eq:kpgg}
\end{eqnarray}
where $y=p (q_{1}-q_{2})/m_{K}^{2}$ and $z\,=%
\,(q_{1}+q_{2})^{2}/m_{K}^{2}$.
 Then the double differential rate is given
by 
\begin{equation}
\frac{\displaystyle \partial ^{2}\Gamma }{\displaystyle \partial y\,\partial %
z}\sim [%
\,z^{2}\,|\,A\,+\,B\,|^{2}\,+\,\left( y^{2}-\frac{\displaystyle \lambda
(1,r_{\pi }^{2},z)}{\displaystyle 4}\right) ^{2}\,|\,B\,|^{2}\,]~,
\label{eq:doudif}
\end{equation}
where $\lambda (a,b,c)$ is the usual kinematical function 
and $r_{\pi }=m_{%
\pi }/m_{K}$.
 Thus in the region of small $z$ (collinear photons) the $B$
amplitude is dominant and can be determined separately from the $A$
amplitude. This feature is crucial in order to disentangle the CP-conserving
contribution $K_{L}\rightarrow \pi ^{0}e^{+}e^{-}$. In fact  
the lepton pair in 
Fig. \ref{fig:klpee} produced by  photons  in $S$-wave, like 
an ${ A}(z)$-amplitude, are suppressed  by the lepton mass while
the photons  in $B(z,y)$ are
also in $D$-wave and so the resulting  $K_{L}\rightarrow \pi ^{0}e^{+}e^{-}$
 amplitude, 
$A(K_{L}\rightarrow \pi ^{0}e^{+}e^{-})_{CPC}$,  
 does not suffer from  the electron mass suppression \cite{kpee,EPR88}.
\begin{figure}[t]
\begin{center}
\leavevmode
\epsfysize=5 cm
\epsfxsize=10 cm
\epsfbox{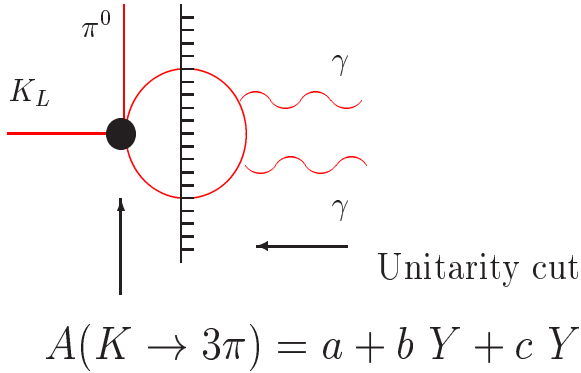}
\end{center}
\caption{$K_L\rightarrow \pi ^0 \gamma \gamma $: 
unitarity contributions from 
 $K \rightarrow 3 \pi$: $X,Y$ Dalitz variables \cite{CD93,CE93}}
\label{fig:klpggunit} 
\end{figure}

 The leading  $\mathcal{O}(p^{4})$ $K_L\rightarrow \pi ^0 \gamma \gamma$
amplitude \cite{Cappiello}
 is affected by two  large  $\mathcal{O}(p^{6})$ contributions: 
i) the full unitarity corrections from $K \rightarrow 3 \pi$
 \cite{CD93,CE93} in Fig. \ref{fig:klpggunit}  and   
ii) local contributions.
Fig. \ref{fig:klpggunit}  
enhances the $\mathcal{O}(p^{4})$ branching ratio
 by 
$40\%$ and generates a $B$-type amplitude.
 At this order there are 
three independent counterterms, as the one in Eq. (\ref{eq:Ksp6}),
 with the unknown coefficients 
$\alpha_1, \alpha_2$ 
and $\beta$ leading to contributions to $A$ and $B$  
in Eq. (\ref{eq:kpgg}) \cite{CE93}:
\begin{equation}
A_{\rm CT}=\alpha_1 (z-r_\pi^2)+\alpha_2, \quad
 B_{\rm CT}=\beta 
\label{eq:AB}.
\end{equation}
\begin{figure}[t]
\begin{center}
\leavevmode
\epsfysize=7.5 cm
\epsfxsize=10 cm
\epsfbox{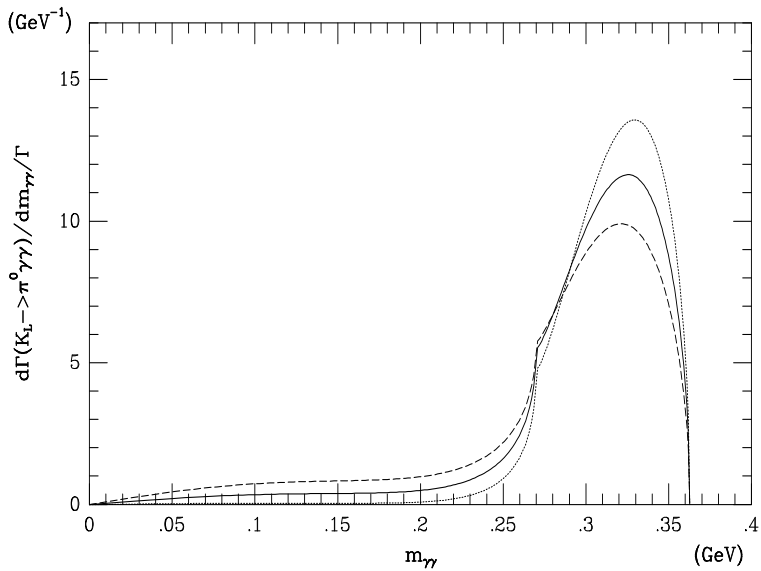}
\end{center}
\label{fig:klpggspectrum}
\caption{$K_L\rightarrow\pi^0 \gamma\gamma$ diphoton-invariant mass
spectrum for different values of $a_V$ : $-$0.4 (dotted curve),  
$-$0.7 (full curve), $-$1 (dashed curve) \cite{DP97}.} 
\end{figure}
If we assume 
  VMD \cite{SE88,EP90},
these couplings are related in terms of one constant, $a_V$:
\begin{equation}
\alpha_1=\frac{\beta}{2}=-\frac{\alpha_2}{3}=-4a_V.
\end{equation}
Though chiral counting suggests $\alpha_i ,\beta\sim 0.2 $,
VMD enhances this typical size. 
Actually a model, FMV, 
describing weak interactions of 
pseudoscalars ($\phi$'s)  
with vectors,
 ${\cal L}_W ^{FMV} (\phi,V^\mu)$,
based on factorization and couplings fixed by the Wilson coefficient
of the $Q_{-}$ operator,
 predicts \footnote{Due to the better ultraviolet behaviour, as discussed in 
Section \ref{sec:klmumu}, we consider only FMV  contributions for the predictions.}
the size and the sign of the weak VMD couplings \cite{DP97}:
\begin{equation} 
{\cal L}_W ^{FMV} (\phi,V^\mu) \quad \Longrightarrow 
\quad a_V=-0.6. \label{eq:FMV}
\end{equation}
As we can see from Fig. \ref{fig:klpggspectrum} the spectrum at low $z$
is very sensitive to the value of $a_V$, or more generally to the size 
of the amplitude $B$ in Eq. (\ref{eq:kpgg}).
\noindent
Recently Gabbiani and Valencia \cite{GV}  suggested to
fit the experimental 
$z$-spectrum (and the rate) with all three parameters in Eq. (\ref{eq:AB}):
 their preferred solution for 
  $B$ and consequently 
$B(K_L\rightarrow\pi^0 e^+ e^-)_{CPC}$
is very large in fact at low diphoton invariant mass it is   consistent 
with our plot in Fig. \ref{fig:klpggspectrum}
with    $a_V: -1$. The recent data from NA48 \cite{NA48kpgg}
 measure this 
region extremely well  
and exclude this possibility by finding  $a_V : -0.46\pm 0.05$.
\begin{equation}
{\rm NA}48 \quad \Rightarrow \quad B(K_L\rightarrow \pi^0 e^+ e^-)_{CPC}
^{\gamma \gamma \ {\rm on-shell }}\sim 1\times 10^{-12}. 
\end{equation}
As a result we think that the size $B(K_L\rightarrow \pi^0 e^+ e^-)_{CPC}
^{\gamma \gamma \ {\rm on-shell }}$,
is an  issue that can be established firmly from the
$K_L\rightarrow\pi^0 \gamma\gamma$ spectrum. 
More disturbing is the dependence on the form factor when the two 
intermediate photons are off-shell. More theoretical work is needed and
probably a partial answer can come from the measurement of
$K_L\rightarrow\pi^0\gamma\gamma^*$ \cite{kpeeg,Gabbiani97}.

The \underline{$K^+\rightarrow\pi^+\gamma\gamma$}
 channel can be studied in much the same way as
 $K_L\rightarrow\pi^0\gamma\gamma$ and might be  interesting
  for future E949 \cite{Littenberg} and NA48b \cite{NA48KSb}
  experiments. Since the $K^+$ is not 
CP-eigenstate,
in addition to  $A$ and $B$ in eq.(\ref{eq:kpgg}), 
also a helicity amplitude with CP=$-1$ and photons in the $P$-wave
 is allowed but found  small \cite{EPR88}.
The $A$ and $B$ amplitudes receive: i) a $\pi   \pi$-loop contribution
 analogous to 
Fig. \ref{fig:klpggunit} \cite{EPR88,DP96} and ii) $\mathcal{O}(p^{4})$
local contributions from Eq. (\ref{eq:wp4}), $\hat{c}$ \cite{EPR88}, and
small $\mathcal{O}(p^{6})$ VMD contributions \cite{DP96}.
 BNL787 with 31 events
has measured
$B(K^+\rightarrow\pi^+\gamma\gamma)=(6\pm 1.6)\times 10^{-7}$ and 
$\hat{c}=(1.8\pm 0.6)$ \cite{BNL787},
which has interesting dynamical implications \cite{EKW93,DP98}.

\section{$K^\pm \rightarrow \pi ^{\pm} 
\ell ^{+}\ell ^{-}$ and $K_{S} \rightarrow
\pi ^{0}\ell ^{+}\ell ^{-}$ }
The CP-conserving decays $K^\pm (K_{S}) \rightarrow \pi ^{\pm} 
(\pi ^{0})\ell ^{+}\ell ^{-}$ are dominated by the 
long-distance process $K\to\pi\gamma \to \pi   \ell^+ \ell^-$ \cite{EPR}.
The decay amplitudes can in general be written in terms of one form 
factor $W_i(z)$ ($i=\pm,S$):
\begin{equation}
A\left( K_i \rightarrow \pi ^i \ell^+ \ell^- \right) = - \frac{%
\displaystyle e^2}{\displaystyle M_K^2 (4 \pi)^2} W_i(z) (k+p)^\mu \bar{u}%
_\ell(p_-) \gamma_\mu v_\ell(p_+)~,\label{eq:CPV_S}
\end{equation}
$z=q^2/ M_K^2$; $W_i(z)$  can be decomposed as the sum of 
a polynomial piece plus a
non-analytic term, $W_{i}^{\pi \pi}(z)$, generated  by  
the $\pi \pi $ loop, analogously  to the one in Fig. \ref{fig:klpggunit} 
for $K_L\rightarrow\pi^0\gamma\gamma$,  completely determined in terms 
of the physical $K\rightarrow 3 \pi$ amplitude \cite{DEIP}.
Keeping the polynomial terms up to  $\mathcal{O(}p^{6})$
we can write 
\begin{equation}
W_{i}(z)\,=\,G_{F}M_{K}^{2}\,(a_{i}\,+\,b_{i}z)\,+\,W_{i}^{\pi \pi
}(z)\;,
\label{eq:Wp6}
\end{equation}
where the  parameters $a_{i}$ and $b_{i}$ parametrize local 
contributions starting respectively at $\mathcal{O(}p^{4})$  
and $\mathcal{O(}p^{6})$. 
Recent data on $K^{+}\rightarrow \pi ^{+}e^{+}e^{-}$ and
$K^{+}\rightarrow \pi ^{+}\mu ^{+}\mu ^{-}$  by BNL-E865
\cite{kpllE865} have been successfully\footnote{For 
an alternative description of the data see \cite{burk00}.}
fitted using 
Eq. (\ref{eq:Wp6})
and lead to
\begin{equation}
a_{+}\,=-0.587\pm 0.010,\qquad \,b_{+}=-0.655\pm 0.044~.  
\label{eq:ab+}
\end{equation}

Recentely HyperCP \cite{HyperCP} has measured  
the CP-violating width  charge asymmetry  in
 $K^{\pm}\rightarrow \pi ^{\pm} \mu ^{+}\mu ^{-}$ and 
it has found that it is consistent with 0
at 10\% level. Though the CKM prediction with accurate cuts
 is $\sim 10^{-4}$
 \cite{DEIP}, we are beginning to test new physics
affecting the operator $\bar{s} d\bar{\mu} \mu$. 
The experimental size of the ratio $b_{+}/a_{+}$
exceeds the naive dimensional analysis estimate
$b_{+}/a_{+} \sim {\cal O}[M^2_K/(4\pi F_\pi)^2] \sim 0.2$, but can 
 be explained by a large VMD contribution.
Chiral symmetry alone does not allow  us to 
determine the unknown couplings $a_{S}$ and $b_S$ 
in terms of $a_{+}$ and $b_{+}$ \cite{EPR,DEIP}. Neglecting 
 the $\Delta I=3/2$ suppressed  
 non-analytic term $W_{S}^{\pi \pi}(z)$,  we obtain \cite{DEIP}
\begin{equation}
B(K_S \rightarrow \pi^0 e^+ e^-)  =  
\left[46.5 a_S^2 + 12.9 a_S b_S + 1.44 b_S^2 
\right]
\times 10^{-10} 
   \approx     5 \times 10^{-9} \times a_S^2~,
\label{eq:BRKS}
\end{equation}
The recent experimental 
information $B(K_S \rightarrow \pi^0 e^+ e^-) < 1.4 \times 10^{-7}$ 
\cite{NA48KS} let us derive the 
bound $|a_S| \le  5.3$; NA48 \cite{NA48KSb} 
and maybe KLOE \cite{KLOE} will assess in the near future the value 
of this branching at the least for values of $a_S$ of order 1. 
Of course even a strong bound is relevant, since it
 will ensure that this contribution is not dangerous  to measure direct CP
violation in
$K_{L}\rightarrow \pi ^{0}e^{+}e^{-}$. 
We remark that even a sizeable  $a_S$: 
 $a_S<-0.5$ or $a_S>1$, will lead to an interesting interference:
\begin{equation}
B(K_{L}\rightarrow \pi ^{0}e^{+}e^{-})_{CPV}\,=\,\left[
15.3\,a_{S}^{2}\,-\,6.8\frac{\displaystyle \Im \lambda _{t}}{\displaystyle %
10^{-4}}\,a_{S}\,+\,2.8\left( \frac{\displaystyle \Im \lambda _{t}}{%
\displaystyle 10^{-4}}\right) ^{2}\right] \times 10^{-12}~,
\label{eq:cpvtot}
\end{equation}
where $\lambda _{t}= V_{td} V_{ts}$.
\begin{figure}[htb]
\vfill
\begin{minipage}[b]{5.5cm}\centering
\mbox{\epsfig{file=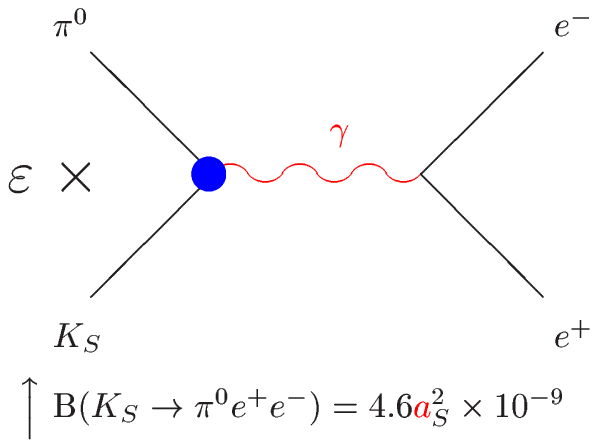,width=5cm, height=3.5cm}}
\caption{Indirect CP violation contribution to
$B(K_{L}\rightarrow \pi ^{0}e^{+}e^{-})$}
\end{minipage}
\hspace{0.8cm}
\begin{minipage}[b]{5.5cm}\centering 
\mbox{\epsfig{file=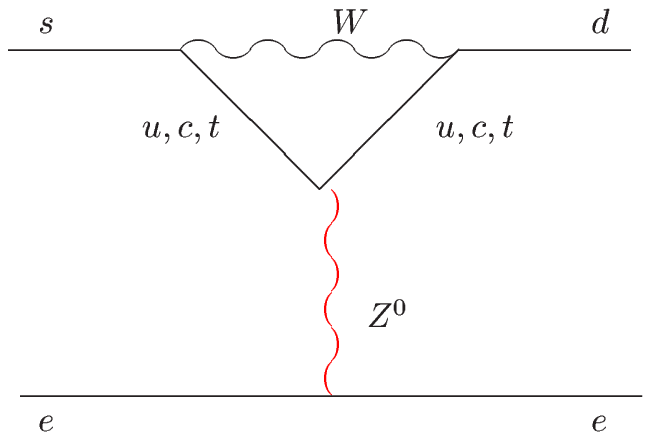,width=5.5cm, height=3.5cm}} 
\caption{Typical direct CP violation contribution to
$B(K_{L}\rightarrow \pi ^{0}e^{+}e^{-})$}
\end{minipage}
\vfill
\end{figure}
The sign of the interference term is model-dependent, but, however is not 
a problem
to determine $\Im \lambda _{t}$ accurately   (up to a discrete
 ambiguity). 

\section{$K\to \pi \pi \gamma $}

We can decompose
$K(p)\rightarrow \pi (p_{1})\pi (p_{2})\gamma (q)$ amplitudes,
according to gauge and Lorentz invariance,
in electric ($E)$ and magnetic ($M)$ terms \cite{DI98,
DMS93,ecker94} 
\begin{equation}
A(K\to \pi \pi \gamma )=\varepsilon _{\mu }(q)\left[ E(z_{i})(p_{1}\cdot
q\,p_{2}^{\mu }-p_{2}\cdot q\,p_{1}^{\mu })+M(z_{i})\epsilon ^{\mu \nu \rho
\sigma }p_{1\nu }p_{2\rho }q_{\sigma }\right] /m_{K}^{3},  \label{amplitude}
\end{equation}
where $z_{i}={p_{i}\cdot q}/m_{K}^{2}$ and $z_{3}=p_{_{K}}\cdot q /{m_{K}^{2}}$.
In the electric transitions one generally separates the bremsstrahlung
amplitude $E_{B}$, theoretically predicted  firmly  by the Low theorem  in 
terms of the non-radiative amplitude and enhanced by the $1/E_\gamma$ 
behaviour.
Summing over photon
helicities, there is no interference among electric and magnetic terms: 
${{\mbox{d} ^2\Gamma }/({\mbox{d}z_{1}\mbox{d}z_{2}}}) \sim 
|E(z_{i})|^{2}+|M(z_{i})|^{2}$. At the lowest order, ($p^{2})$,  one
obtains only $E_{B}$.
Magnetic and electric direct emission amplitudes, 
appearing at ${\cal O} (p^{4}),$
 can be decomposed in a multipole expansion (see Refs.\cite
{DI98,DMS93,LV88,ecker94}). In the table
 below we show the present  experimental status 
with the reason for the suppression
of the bremsstrahlung amplitude and the leading multipoles.

\noindent 
 $
\begin{array}[t]{|c|c|c|}
\hline
\mbox{Decay} & BR(\mbox{bremsstrahlung}) & BR(\mbox{direct 
emission}) \\ \hline
\begin{array}{l}
K_{S}\to \pi ^{+}\pi ^{-}\gamma  \\ 
E_{\gamma }^{*}>50{\rm MeV}
\end{array}
& (1.78\pm 0.05)\times 10^{-3} & <6\times 10^{-5}(E1) \\ \hline
\begin{array}{l}
K_{L}\to \pi ^{+}\pi ^{-}\gamma  \\ 
E_{\gamma }^{*}>20{\rm MeV}
\end{array}
&
\begin{array}{c}
(1.49\pm 0.08)\times 10^{-5} \\ 
({\rm CP}\mathrm{\ violation)}
\end{array}
& 
\begin{array}{c}
(3.09\pm 0.06)\times 10^{-5} \\ 
M1,E2
\end{array} \\ \hline
\begin{array}{l}
K^{\pm }\to \pi ^{\pm }\pi ^{0}\gamma  \\ 
T_{\pi ^{+}}^{*}=(55-90){\rm MeV}
\end{array}
& 
\begin{array}{c}
(2.57\pm 0.16)\times 10^{-4} \\ 
(\Delta I=3/2)
\end{array}
& 
\begin{array}{c}
(4.72\pm 0.77)\times 10^{-6} \\ 
E1,M1
\end{array}
\\ \hline
\end{array}
$

\vspace{0.5cm}
\underline{$K_{S}\rightarrow \pi ^{+}\pi ^{-}\gamma $}.
This channel might be interesting for KLOE \cite{KLOE} and
 NA48 \cite{NA48KSb}: 
only for large
 photon energy might the dynamical interesting  $E1 $--$E_B $ 
interference  be
 observed over the pure bremsstrahlung rate \cite{DMS}. 
The  relevant ${\cal O} (p^4)$ 
counterterm combination to $E1 $  in this channel
is related by chiral symmetry  
  to the one contributing to 
$K^{+}\rightarrow \pi ^{+}\pi ^{0}\gamma $.

\underline{$K_{L}\rightarrow \pi ^{+}\pi ^{-}\gamma $}.
Bremsstrahlung ($E_{B})$ is suppressed by CP violation
 but
enhanced by the $1/E_\gamma $ behaviour. 
KTeV has also measured the magnetic transition M1 with a non-trivial
 form 
factor \cite{KTeV-kppg}:
\begin{equation}
M1=\widetilde{g}_{M1}\left[ 
1+\frac{a}{1-M_{K}^{2}/M_{\rho
}^{2}+2M_{K}E_{\gamma }^{*}/M_{\rho
}^{2}}\right]
\label{amplitudeKLpipig}
\end{equation}
determining $a=(-1.243\pm 0.057$) and the branching
given in the table, which fixes also $\widetilde{g}_{M1}$. 
Interestingly they find that this parametrization is substantially
 better
than a linear fit showing that VMD is at work. In terms of the basic
 ${\cal O} (p^4)$ weak lagrangian in (\ref{eq:wp4}) $M1$ is written as 
\begin{equation}
M1=N_{29}+N_{31}+{\rm h.o.}
\label{amplitudeKLpipigth}
\end{equation}
There are two ways of  implementing VMD in (\ref{eq:wp4})
with different results for $M1$ \cite{EKW93,DP98}. 
KTeV data in (\ref{amplitudeKLpipig}) have shown that
there are large VMD contributions to (\ref{amplitudeKLpipigth}) and 
so  (analogously to the
strong sector), data prefer that VMD be realized  at 
 $\mathcal{O} (p^4)$
\cite{DP98,gao00},
and not  ${\cal O} (p^6)$ \cite{ENP94}.

\underline{$K_{L,S}\rightarrow \pi ^{+}\pi ^{-} e^+ e^- $}
KTeV and NA48 \cite{KTeV-kppee,NA48kpgg} 
have  recently measured 
the asymmetry in the angle between the 
$e^+ \-  e^-$  and the $\pi ^{+} \- \pi ^{-}$ planes
in the decay 
$K_{L}\rightarrow \pi ^{+}\pi ^{-} e^+ e^- $:
this measures the  CP violating interference of the
bremsstrahlung ($E_{B})$ with the $M1$ transition of
$K_{L}\rightarrow \pi ^{+}\pi ^{-}\gamma$ decays, 
enhanced by 
the CP-suppressed
denominator $\Gamma(K_{L}\rightarrow \pi ^{+}\pi ^{-} e^+ e^-) $ ($\sim 
E_{B} ^2)$ .
This quantity is thus very well predicted
in terms of known long-distance observables,
 but it is  not an efficient CKM test \cite{Savage}. 
Recently NA48 has  measured the CP-even bremsstrahlung
dominated decay 
$K_{S}\rightarrow \pi ^{+}\pi ^{-} e^+ e^- $ \cite{NA48-Ksppee}.
The asymmetry in the angle between the 
$e^+ \-  e^-$  and the $\pi ^{+} \- \pi ^{-}$ planes 
for $K_{S}\rightarrow \pi ^{+}\pi ^{-} e^+ e^- $
is small; however, it might  be interesting for NA48 and KLOE 
to test different observables \cite{bulanov}.

\underline{$K^{+}\rightarrow \pi ^{+}\pi ^{0}\gamma $}.
Due to the $\Delta I=3/2$ suppression of the bremss\-trahlung,   
in\-ter\-fer\-ence between $E_B$ and $E1$  and magnetic tran\-si\-tions
can be measured.
New data from BNL\ E787 \cite{E787-00} show van\-ish\-ing
in\-ter\-fer\-ence, 
thus  putting 
a non-trivial bound on model predictions for the counterterm coefficient
in (\ref{eq:wp4}) contributing to $E1$ \cite{EKW93,DP98}.
Consequently  
the direct emission branching ($B(K^{+}\rightarrow \pi ^{+}\pi
^{0}\gamma $)$_{\mathrm{exp}}^{DE}),$ in the table, must be 
interpreted as a
pure magnetic transition and related to the analogous one in 
$K_{L}\rightarrow \pi ^{+}\pi ^{-}\gamma$ \cite{gao00}.

\underline{Direct CP violation}
Direct CP violation can be established in the width charge asymmetry 
in $K^{\pm }\rightarrow \pi ^{\pm }\pi ^{0}\gamma ,$ 
$\delta \Gamma /2\Gamma $,
and \ in the interference $E_{B}$ with $E{1}$ in $K_{L}\rightarrow \pi
^{+}\pi ^{-}\gamma $ ($E1$ with $M_{1}$ in $K_{L}\rightarrow \pi ^{+}\pi
^{-}e^{+}e^{-})$; both observables are kinematically difficult since one is
looking 
in the Dalitz plot
at large photon energy \cite{DI98}. SM charge asymmetries
were looked in \cite{CPold} expecting $\delta \Gamma /2\Gamma \leq 10^{-5}$.
Supersymmetry may enhance this asymmetry  by a factor of 10 \cite{CIP}.

\section{$K_{L}\rightarrow l^{+}l^{-}$}\label{sec:klmumu}

\begin{figure}[htb]
\vfill
\begin{minipage}[b]{4.5cm}\centering
\mbox{\epsfig{file=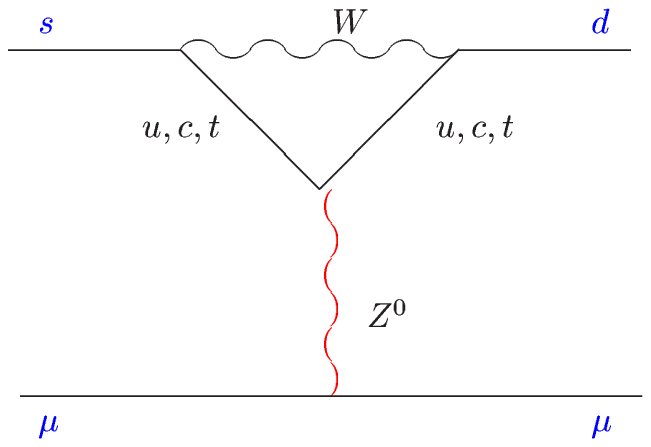, width=4.5cm,height=3.5cm}}
\caption{$A_{\rm SD}$}
\label{s2mumushort}
\end{minipage}
\hspace{0.6cm}
\raisebox{2.4cm}{\parbox[b]{1.5cm}{\Large $<<$}}
\begin{minipage}[b]{5.5cm}
\mbox{\epsfig{file=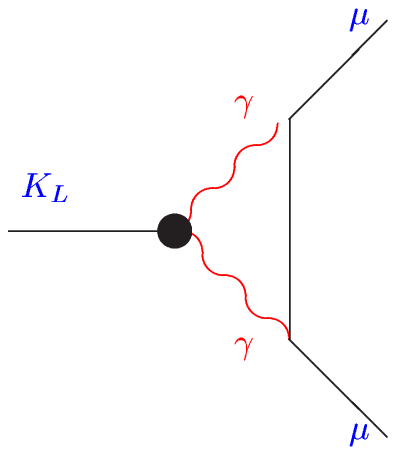,width=5cm,height=3.5cm}}
\caption{$A_{\rm LD}$}
\label{s2mumulong}
\end{minipage}
\vfill
\end{figure}
 

$K_{L}\rightarrow \mu ^{+}\mu ^{-}$
is 
an interesting channel 
to determine $V_{td}$
and to probe new physics.
In the Standard Model the short-distance
contribution, $A_{\rm SD}$, 
is generated by diagrams like the one in Fig. \ref{s2mumushort}.
The known  experimental rate
$\Gamma(K_L\rightarrow
\gamma\gamma)$ \cite{PDG2000} allows us to determine 
  the two-photon 
absorptive contribution, $|\Im  A|^2$,  in Fig. 
\ref{s2mumulong} \cite{seghal69}. This almost
saturates the experimental $K_{L}\rightarrow \mu ^{+}\mu ^{-}$  
rate from E871 \cite{PDG2000,E871klmumu}:
\noindent
\begin{equation}
 {\rm Br}(K_L\rightarrow\mu\overline{\mu})={|\Re A|^2 \hspace{0.25cm}+ 
\hspace{0.25cm}
|\Im A|^2}\stackrel{\rm E871}{=}{(7.18 \pm0.17)}\times 10^{-9}
\end{equation}
\noindent

$\hspace{3cm}
{\swarrow} \hspace{0.6cm} ^{B(K_L\rightarrow
\gamma\gamma)}\big{\downarrow}
\hspace{1cm}$

\noindent
\begin{equation}
\Re ({A_{\rm SD}}+A_{\rm LD}),~~~~ |\Im A|^2=(7.1\pm0.2)\times 10^{-9}
\label{klmumu}
\end{equation}
Thus the  sum of the real parts, long and 
short distance,
$\Re ({A})=  \Re ({A_{\rm SD}}+A_{\rm LD})$, is bound to be very small:
$|\Re ({A}_{\rm exp})|^2 < 4.0 \times 10^{-10} $ 
at $90\%$ C.L.
The known $V_{td}$-dependence of the   SM  short-distance amplitude
 ${A_{\rm SD}}$ \cite{buras01,BB} allows us to obtain 
the bound on 
$\bar{\rho}=\rho(1-\lambda ^2/2)$ \cite{DIP}:
\begin{equation}
\bar{\rho}>1.2-\max \left[ \frac{|\Re A_{\rm exp}|+|\Re A_{\rm LD}|}
{3 \times 10^{-5}}\left( \frac{\bar{m} _t(m_t)}{170 {\rm GeV}}\right)^{-1.55}
\left( \frac{|V_{cb}|}{0.040}\right)^{-2}\right]
\label{eq:rhobound}
\end{equation}
To do better  and constrain  new physics  
it is necessary to have a reliable control
on the model-dependent long-distance dispersive amplitude
$A_{\rm LD}$
in Fig. \ref{s2mumulong} \cite{DIP}-\cite{gomez}. 
In practice one has to understand the proper 
$K_L \rightarrow \gamma ^* \gamma^* $  
form factor, 
$f(q_{1}^{2},q_{2}^{2})$, for Fig. \ref{s2mumulong}. 
In Ref. \cite{gomez}, in analogy to the real 
decay  $K_L \rightarrow \gamma  \gamma $,
this  has  first been written as the 
 sum of the poles $\pi ^0, \eta$ and $\eta '$;  then weak couplings are 
determined by a 
 large-$N_{c}$ argument and $U(3)\otimes U(3)$ symmetry, while 
the experimental knowledge of the electromagnetic decays of pseudoscalars, 
$P$,
$P\rightarrow e^+ e^-$ constrains the relevant local contribution. 
Somehow the 
problem of the form factor has been thus bypassed. However 
we think the form factor is very sensitive to symmetry breaking
and thus caution must be used before  completely
accepting this result \cite{derafael}.
We instead have proposed \cite{DIP} a 
low energy parameterization of the 
$K_{L}\rightarrow \gamma ^{*}\gamma ^{*}$ form factor that 
includes the poles
of the lowest vector meson resonances ($m_{V}\sim m_{\rho}$): 
\begin{equation}
f(q_{1}^{2},q_{2}^{2})=1+\,\alpha \left( \frac{\displaystyle q_{1}^{2}}{%
\displaystyle q_{1}^{2}-m_{V}^{2}}+\frac{\displaystyle q_{2}^{2}}{%
\displaystyle q_{2}^{2}-m_{V}^{2}}\right) \,+\,\beta \,
\frac{\displaystyle %
q_{1}^{2}q_{2}^{2}}{\displaystyle (q_{1}^{2}-m_{V}^{2})
(q_{2}^{2}-m_{V}^{2})}~.  \label{eq:fq1q2}
\end{equation}
The ansatz is that, since we are able to describe
 the relevant resonances fully, 
this is the proper form factor
to high energy up to the charm scale and in fact
by  comparing it with the short-distance result in \cite{BB}
 we constrain the parameters $\alpha $ and $\beta $:
the form factor in Eq.~(\ref
{eq:fq1q2}) goes as $1+2\alpha +\beta $ for $q_{i}^{2}\gg m_{V}^{2}$ 
and thus
 the logarithmically divergent 
$A_{LD}$ in  Fig. \ref{s2mumulong} 
can be phenomenologically compared with the known 
perturbative QCD calculation \cite{BB}
leading to  $|1+2\alpha +\beta |\ln (\Lambda /m_{V})<0.4$ ($\Lambda$ is 
the  ultraviolet cutoff) and limiting $\beta $
for a fixed value of $\alpha $.
There are two important questions that we will address now to 
 establish  $A_{SD}$ accurately and safely: i) the  
experimental or theoretical determination
of the parameters $\alpha $ and $\beta $ 
(expected to be ${\cal O}(1)$ by
chiral power counting)  
and ii) making  sure that the form factor in Eq.~(\ref
{eq:fq1q2}) is correct.
We can expand Eq. (\ref{eq:fq1q2})
 for $q_{2}^{2}=0$ and 
$q_{1}^{2}\ll m_V^2$, obtaining
\begin{equation}
f(q_{1}^{2} \ll m_V^2,q_{2}^{2}=0)
=1-\,0.42\alpha \frac{\displaystyle q_{1}^{2}}{\displaystyle m_K^2}
-\,0.17\alpha \frac{\displaystyle q_{1}^{4}}{\displaystyle m_K^4}+
{\rm h.o.}
 \label{eq:fdipex}
\end{equation}
and describe simultaneously
$K_{L}\rightarrow e ^{+} e ^{-} \gamma  $ and
$ K_{L}\rightarrow  \mu ^{+}\mu ^{-} \gamma$  decays. 
However, data  are not yet sufficient  to clearly show  if
Eq. (\ref{eq:fdipex}) is a better description of these decays than, 
for example, a linear fit. Historically
data have been analysed using simply the BMS form factor \cite{BMS},
this
is still VMD motivated and,   in the low energy region it is similar to
the form factor (\ref{eq:fdipex}), but for $q^{2}\gg  m_V^2$ cannot 
match
QCD and thus must be regarded as  a low energy phenomenological model. 
The low energy parametrization of the  BMS model is: 
\begin{equation}
f_{BMS}(q_{1}^{2} \ll m_V^2)
=1+(0.42-1.3 \alpha _{K^*})\frac{\displaystyle q_{1}^{2}}{\displaystyle m_K^2}
+(0.17-0.91 \alpha _{K^*}) \frac{\displaystyle q_{1}^{4}}
{\displaystyle m_K^4}+
{\rm h.o.}
\label{eq:fbmsex}
\end{equation}

In fact $K_{L}\rightarrow e ^{+}e ^{-} \gamma  $ \cite{NA48kleeg} 
has been analysed using only
(\ref{eq:fbmsex}) and finding $\alpha _{K^*}=(-0.36\pm 0.1)$. I have checked
that (\ref{eq:fdipex}) with $\alpha =-1.5$ fits even better the
$K_{L}\rightarrow e ^{+} e ^{-} \gamma  $ spectrum.
 KTeV has recently measured the 
$ K_{L}\rightarrow  \mu ^{+}\mu ^{-} \gamma$ 
spectrum and rate  with (\ref{eq:fdipex})  and (\ref{eq:fbmsex}),
 finding respectively $\alpha =-1.54 \pm 0.10 $ 
and $\alpha _{K^*} =-0.160^{+0.026} _{-0.028} $ \cite{KTeVklmumug}: 
for these values
even the  quadratic
slopes in (\ref{eq:fdipex})  and (\ref{eq:fbmsex}) agree. 
However it seems that
 the BMS model does not fit simultaneously
$K_{L}\rightarrow e ^{+} e ^{-} \gamma  $ 
and $ K_{L}\rightarrow  \mu ^{+}\mu ^{-} \gamma$ 
spectra but this could be also  caused by some experimental problem.
We look forward to a clear determination of the linear and quadratic 
slopes
in both lepton channels,
so to clearly establish  that the form factors 
in (\ref{eq:fdipex})  and (\ref{eq:fbmsex}) are better than the 
linear slope.
For the values  in Ref.\cite{KTeVklmumug} 
the difference is marginal.
We stress that the advantage of our model is the good 
 behaviour for large $q^{2}$.
Another important test is the measurement of the quadratic slope $ 
\beta $ in
(\ref{eq:fq1q2}) from
$K_{L}\rightarrow e ^{+} e ^{-} \mu ^{+}\mu ^{-}$ \cite{KTeVkleemumu} or 
$K_{L}\rightarrow e ^{+} e ^{-} e ^{+}e ^{-}$ \cite{KTeVkleeee,NA48kleeee}. 
Of course this is a difficult measurement;
 however, encouraging results
have been obtained lately in  $e ^{+} e ^{-} \mu ^{+}\mu ^{-}$
(43 events)  by KTeV \cite{KTeVkleemumu}
and $e ^{+} e ^{-} e ^{+}e ^{-}$ (441 events) 
 by KTeV \cite{KTeVkleeee} and by NA48 \cite{NA48kleeee}, 
where the branchings have been obtained 
and although  $ \beta $ is not determined yet, the measurement of 
the form factor is
not so far away since the linear terms have  already been studied. 
\cite{KTeVkleeee}

 Now, if we take:  $\beta $ from the matching conditions and
 the latest experimental determinations of 
$\Gamma (K_{L}\rightarrow \mu ^{+}\mu ^{-})$,
$\Gamma (K_{L}\rightarrow \gamma \gamma )$ and $\alpha$, we obtain   
 $|{\Re {\cal A}}_{LD}|<2.07\times 
10^{-5}$
and $\rho >-0.2$ at $90\%$ $C.L.$ \cite{KTeVklmumug,gino01}.
This bound could be  improved and made more solid if the form factor in
 Eq.~(\ref
{eq:fq1q2}) were firmly  established and the parameters
were measured with good precision.  An encouraging   result is
also that the experimental
value for $\alpha$ follows the 
theoretical prediction in Ref.\cite{DP97} (see also  Eq. (\ref{eq:FMV})):
\begin{equation}
{\cal L}_W ^{FMV}(\phi,V^\mu)\Longrightarrow \alpha=1.2
\end{equation}
based on short distance,
showing that the low energy description in
 Eq.~(\ref{eq:fq1q2}) is able to capture also short-distance physics.
Also lattice \cite{marti} might help to establish the correct 
form factor.
Recently $K_{L}\rightarrow e^{+}e^{-}$
 has been measured at BNL by E871 as \cite{E871klee}:
 $B(K_{L}\rightarrow e^{+}e^{-})=(8.7_{-4.1}^{+5.7})\times 10^{-12}$; 
however,
the theoretical prediction \cite{V98,gomez} for this branching is not
sensitive to the slopes of the form factor but only to $f(0,0)$.

\section{Conclusions}
 I think that we have heard   at this Conference
and I have summarized here some relevant progress:
 the improved measurements of  
$K_{S}\rightarrow  \gamma  \gamma $, $K_{L}\rightarrow \pi ^0
  \gamma  \gamma $,  $K_{L}\rightarrow \pi ^+ \pi ^- \gamma  $ and
$ K_{L}\rightarrow  l ^{+} l ^{-} \gamma (\gamma ^*)$ decays. These 
are useful pieces of  information, which will
serve  to improve our ability  in testing the SM
and understand QCD. Soon we will have interesting data from KLOE,
NA48b and E949 \cite{NA48kpgg,Littenberg,KLOE} 
so that other channels such as
$K_{S}\rightarrow  \pi ^0 e^{+}e^{-}  $, 
$K^+\rightarrow \pi ^+  \gamma \gamma  $ and interesting 
CP-violating asymmetries, e.g. the charge asymmetry in  
$K^+\rightarrow \pi ^+  \pi ^0 \gamma$ and
 $K^+\rightarrow \pi ^+  \mu ^+\mu^-$ \cite{HyperCP},
 will be measured. We have seen that 
our ability to test the SM in
$K_{L}\rightarrow \pi ^0 e ^{+} e ^{-}$ and 
$ K_{L}\rightarrow  \mu ^{+} \mu ^{-}$  depends crucially  
on how good we match short distance: 
here theoretical progress has been made \cite{derafael}
and more is needed. 
Other interesting prospects are the muon polarization in
$K_L\rightarrow \pi ^0  \mu ^+\mu^-$\cite{diwan} and 
time interferences in 
$K_{L,S}\rightarrow \pi ^0 e ^{+} e ^{-}$ \cite{belyaev} to definitely
suppress Greenlee background.

\section*{Acknowledgements}
I would like to thank the organizers for the invitation to this
stimulating conference, Gino Isidori and Daoneng Gao
for nice discussions and for reading the manuscript. Also I thank
Gerhard Buchalla
for important
collaboration  and CERN-TH division for nice hospitality.
This work was supported in part by
TMR, EC--Contract No. ERBFMRX-CT980169
(EURODA$\Phi$NE).


\begin{thebibliography}{99}

\bibitem{buchalla}  G.\ Buchalla, hep-ph/0002207; these Proceedings,
hep-ph/0110313 and references therein.

\bibitem{buras01} A.J. Buras, hep-ph/0109197; these Proceedings,
 hep-ph/0109197 and references therein.

\bibitem{NP} A. Masiero, these Proceedings; I. Bigi, these  Proceedings,
hep-ph/0107102;
L. Silvestrini, these Proceedings, 
hep-ph/0107102;   G. D'Ambrosio and Dao-Neng Gao, 
Phys. Lett.  {\bf B 513} 123 (2001), hep-ph/0105078.

\bibitem{NA48kpgg}  M.Martini, NA48 collaboration, these Proceedings;
L Iconomidou-Fayard,  NA48 Collaboration, hep-ex/0110028.
 
\bibitem{barker00} A.R. Barker and S.H. Kettell, Ann. Rev. Nucl. Part. Sci.
{\bf 50} 249 (2000), hep-ex/0009024. 

\bibitem{Littenberg} L. Littenberg, these Proceedings and 
hep-ex/0010048.

\bibitem{HyperCP}H. K. Park {\it et al.}, 
HyperCP collaboration, hep-ex/0110033; M.J. Longo, these Proceedings.

\bibitem{KLOE} T. Spadaro, KLOE collaboration, these Proceedings.

\bibitem{NA48KSb} M. Calvetti,  these Proceedings.

\bibitem{Weinberg1}  S. Weinberg,{ Physica} {\bf A \ 96}, 327 (1979); J.
Gasser and H. Leutwyler, { Ann. Phys.} (N.Y.) {\bf 158, }142 (1984); 
A.V.
Manohar and H. Georgi, { Nucl. Phys.}{\bf \ B } {\bf 234,} 189 (1984).

\bibitem{DI98}  G. D'Ambrosio and G. Isidori, 
Int. J. Mod. Phys.\textit{\ } 
\textbf{A 13}, {1} {(1998). } 

\bibitem{Eckerp} G. Ecker, these Proceedings,
hep-ph/0108231 and references therein; 
J. Bijnens, hep-ph/0108111.


\bibitem{EKW93} G. Ecker, J. Kambor and D. Wyler, { Nucl. Phys.} B
{\bf \
394,} 101 (1993).

\bibitem{DP98}   G. D'Ambrosio and J. Portol{\'e}s, 
Nucl. Phys. \textbf{B 533,}
494 (1998).

\bibitem{derafael}  
M.\ Knecht, S.\ Peris, M.\ Perrotet and E. de Rafael, 
Phys. Rev. Lett. \textbf{83}, 5230 (1999); E. de Rafael, 
these Proceedings, hep-ph/0109280; J.F.\  Donoghue,  these Proceedings.


\bibitem{GL}  M.K. Gaillard and B.W. Lee, { Phys. Rev. Lett.} {\bf 33},  
 108 (1974).

\bibitem{DEG}  G. D'Ambrosio and D. Espriu, Phys. Lett. \textbf{B
175}, 237 (1986); J.L. Goity, Z. Phys\emph{.} \textbf{C 34,} {341}
 {(1987)}.


\bibitem{NA48Ksgg} A. Lai {\it et al.},  NA48 Collaboration, 
Phys. Lett. \textbf{B 493} 29 (2000). 


\bibitem{KH94}  J. Kambor and B.R. Holstein, Phys. Rev. \textbf{D 49, }2346
(1994).

\bibitem{gino98a} G. Buchalla and G.\ Isidori, Phys. Lett. \textbf{B 440,}
170 (1998) and references therein.

\bibitem{DG95}  J.F. Donoghue and F. Gabbiani, Phys. Rev. \textbf{D 51,} 2187
(1995).

\bibitem{greenlee}  H.B.\ Greenlee, Phys. Rev. \textbf{D 42, }3724 (1990).

\bibitem{KTeVklpi0ll}  A. Alavi-Harati \textit{et al.}, KTeV Collaboration,
Phys. Rev. Lett. \textbf{84}, 5279 (2000); A. Alavi-Harati \textit{et al.},
hep-ex/0009030.

\bibitem{diwan} Diwan, M.L. {\it et al}, these Proceedings, hep-ex/0108025.

\bibitem{kpee} J.F. Donoghue, B.\ Holstein and G.\ Valencia, Phys. Rev. 
\textbf{D 35, }2769 (1986);
T. Morozumi and H. Iwasaki, Prog. Theor. Phys. \textbf{82}, 371 (1989);
  J. Flynn and L. Randall, Phys. Lett. \textbf{B
216} 221 (1989). 

\bibitem{EPR88} {G.} Ecker, A. Pich and E. de Rafael, Nucl.
Phys. \textbf{B 303, }665 (1988). 

\bibitem{Cappiello}  G. Ecker, A. Pich and E. de Rafael, Phys. Lett. \textbf{
B 189}, {363} {(1987)}; L. Cappiello and G. D'Ambrosio, Nuovo Cim.\textit{\ } 
\textbf{A 99}, 155 (1988).

\bibitem{SE88}    L.M. Sehgal, Phys. Rev\emph{.} \textbf{D 38,} 808 (1988);
P.\ Heiliger and L.M. Sehgal, Phys. Rev\emph{.} \textbf{D 47,\ }4920 (1993).


\bibitem{CD93}  L. Cappiello, G. D'Ambrosio and M. Miragliuolo, Phys. Lett. 
\textbf{B 298, }423 (1993).

\bibitem{CE93}  A. G. Cohen, G. Ecker and A. Pich, Phys. Lett. \textbf{B
304, } 347 (1993).

\bibitem{EP90}  G. Ecker, A. Pich and E. de Rafael, Phys. Lett. \textbf{B
237,} 481 (1990).

\bibitem{DP97} G. D'Ambrosio and J. Portol{\'e}s, Nucl. Phys.
 \textbf{B 492},\textbf{\ }417 (1997).

\bibitem{GV}
F.~Gabbiani and G.~Valencia,
Phys.\ Rev.  {\bf  D 64}, 094008 (2001), hep-ph/0105006.

\bibitem{kpeeg}  A.~Alavi-Harati {\it et al.},  KTeV Collaboration,
Phys. Rev. Lett.\textbf{83,} 021801 (2001)  
hep-ex/0011093;
K. Murakami \textit{et al.}, KEK E162 Collaboration, Phys. Lett.\textbf{B
463}, 333 (1999).

\bibitem{Gabbiani97}  J.F. Donoghue and F. Gabbiani, Phys. Rev. 
\textbf{D 56,} 1605 (1997).


\bibitem{DP96}  G. D'Ambrosio and J. Portol{\'e}s, Phys. Lett. \textbf{B
389}, 770 (1996).

\bibitem{BNL787} P. Kitching \textit{et al.}, E787 Collaboration, 
Phys. Rev. Lett.
\textbf{79}, 4079 (1997).
 
\bibitem{EPR}  G. Ecker, A. Pich and E. de Rafael, Nucl. Phys. \textbf{B
291,} 692 (1987).

\bibitem{DEIP}  G. D'Ambrosio, G. Ecker, G. Isidori and J. Portol{\'e}s, 
JHEP  \textbf{08,} 004 (1998).


\bibitem{kpllE865}  R.Appel \textit{et al.}, 
E865\ collaboration, Phys. Rev.
Lett. \textbf{83}, 4482 (1999); H.Ma \textit{et al.}, E865\ collaboration,
  Phys.Rev.Lett. \textbf{84} 2580-2583 (2000), hep-ex/9910047. 


\bibitem{burk00} H.Burkhardt \textit{et al.}, Phys. Lett. \textbf{B 512},
317 (2001), hep-ph/0011345. 

\bibitem{NA48KS}
A. Lai {\it et al.},  NA48 Collaboration, 
Phys. Lett.\textbf{B 514}, 253 (2001).

\bibitem{DMS93}  G. D'Ambrosio, M. Miragliuolo and P.\ Santorelli,
``Radiative non-leptonic kaon decays'' in 
\textit{The  DA$\Phi $NE Physics Handbook}, 
Eds. L. Maiani, G. Pancheri and N. Paver, (LNF-Frascati 1992) p. 231.

\bibitem{LV88}  Y.-C.R. Lin and G. Valencia, Phys. Rev. \textbf{D 37,} 143
(1988).

\bibitem{ecker94}  G. Ecker, H. Neufeld and A. Pich,\textbf{\ }Nucl. Phys.%
\textbf{\ B 314, }321 (1994);  G. D'Ambrosio, G. Ecker, G. Isidori and 
H. Neufeld,``Radiative non-leptonic kaon decays'' in 
\textit{The Second DA$\Phi $NE Physics Handbook}, Eds. L. Maiani, G.
Pancheri and N. Paver, (LNF-Frascati, 1995) p.265, hep-ph/9411439.

\bibitem{HS93}  P. Heiliger and L.M. Sehgal,  Phys. Lett. \textbf{B 307, }
{182 (1993)}. 

\bibitem{DMS}  G. D'Ambrosio, M. Miragliuolo and F. Sannino, Z. Phys. 
\textbf{\ C 59,} 451 (1993); G. D'Ambrosio and G. Isidori, Z. Phys. \textbf{C 65,} 649
(1995).

\bibitem{KTeV-kppg}  A.Alavi-Harati {\it et al.} , KTeV
Col\-la\-bo\-ra\-tion, Phys. Rev. Lett. \emph{\ }\textbf{86}, 761 (2001).

\bibitem{gao00}  G.~D'Ambrosio and Dao-Neng Gao, JHEP  \textbf{0010,} 043
 (2000), hep-ph/0010122.

\bibitem{ENP94}  G. Ecker, H. Neufeld and A. Pich, Nucl. Phys.
\textbf{\ B 314,} 321 (1994).

\bibitem{KTeV-kppee} A.Alavi-Harati {\it et al.}, KTeV 
Col\-la\-bo\-ra\-tion, Phys. Rev. Lett. \emph{\ }\textbf{84}, 408 (2000).

\bibitem{Savage}  L.M. Sehgal and M. Wanninger, Phys. Rev\textit{.} 
\textbf{D 46,} 1035 and 5209 (E)(1992);
 ; P. Heiliger and L.M. Sehgal, Phys.
Rev. \textbf{D 48, }4146 (1993); J.K. Elwood, M. Savage and M.B. Wise, Phys.
Rev\textit{.} \textbf{D 52,} {5095 (1995)}; J.K. Elwood  \textit{ibid.} 
\textbf{D 53, }2855\textbf{\ (}E) (1996); J.K. Elwood  \textit{et al}.
Phys. Rev\textit{.} \textbf{D 53,} 4078 (1996);
G. Ecker and H. Pichl, Phys. Lett. \textbf{B 507}, 193 (2001),
hep-ph/0101097.

\bibitem{NA48-Ksppee} A. Lai {\it et al.},  NA48 Collaboration, 
Phys. Lett. \textbf{B 496} 137 (2000). 

\bibitem{bulanov} S.S. Bulanov, hep-ph/0109025.

\bibitem{E787-00} S.C. Adler {\it et al.}, E787 Collaboration,
Phys. Rev. Lett. \textbf{85}, 4856 
(2000).

\bibitem{CPold}  C.O. Dib and R.D. Peccei, Phys. Lett. \textbf{B 249, }{325
(1990)}; N. Paver, Riazzudin and F. Simeoni, Phys. Lett. \textbf{B 316, } {\
397} {(1993);} H.Y. Cheng, Phys. Rev. \textbf{D 49} 3771 (1994).


\bibitem{CIP}  G.\ Colangelo, G.\ Isidori and J. Portol{\'e}s, 
Phys.Lett. \textbf{B 470}, 134 (1999).

\bibitem{PDG2000} Particle Data Group, D.E. Groom {\it et al.},
{Eur. Phys. J. } \textbf{C 15} 514 (2000). 

\bibitem{seghal69} L.M. Sehgal, Phys. Rev. \textbf{ 183},
 {1511} (1969).

\bibitem{E871klmumu} D. Ambrose {\it et al.}, E871 collaboration,
 Phys. Rev. Lett. \textbf{84,}  1389 (2000).

\bibitem{BB}  G. Buchalla and A.J. Buras, {\em Nucl. Phys.}, B
{\bf \ 412} 106
(1994); see also J.O. Eeg, K. Kumericki, I. Picek {Eur. Phys. J. }
\textbf{C 1}  531 (1998).


\bibitem{DIP}  {G. D'Ambrosio, G. Isidori and J. Portol\'{e}s}, {
Phys. Lett.} B {\bf 423}, 385 (1998).

\bibitem{V98}  G. Valencia, { Nucl. Phys.} B{\bf \ 517,} 339 (1998).

\bibitem{gomez}  D. Gomez Dumm and T.\ Pich, { Phys. Rev. Lett.} {\bf 80,}
4633 (1998).

\bibitem{BMS} L. Bergstr${\rm \ddot{o}}$m, E. Masso and P. Singer, 
{  Phys. Lett.} { \bf B 131}, 223 (1983).

\bibitem{NA48kleeg}  V. Fanti \textit{et al.}, NA48 Collaboration,
 Phys. Lett. \textbf{B} {\bf 458,} 553 (1999).

\bibitem{KTeVklmumug} A. Alavi-Harati {\it et al.}, KTeV
Col\-la\-bo\-ra\-tion, Phys. Rev. Lett. {\bf 87}, 071801 (2001).

\bibitem{NA48kleeee} A. Lai {\it et al.}, NA48 Collaboration,
Contributed to 30th International Conference on High-Energy Physics 
(ICHEP 2000), hep-ex/0108037.

\bibitem{KTeVkleeee} A.Alavi-Harati {\it et al.}, KTeV
Col\-la\-bo\-ra\-tion, Phys. Rev. Lett. \textbf{{86}}, 5425 (2001).

\bibitem{KTeVkleemumu} A.Alavi-Harati et al., KTeV
Col\-la\-bo\-ra\-tion, Phys. Rev. Lett. \textbf{87}, 111802 (2001).

\bibitem{gino01} G.Isidori, hep-ph/0110255.

\bibitem{marti}G. Martinelli, these Proceedings, hep-ph/0110023. 

\bibitem{E871klee} D. Ambrose {\it et al.}, E871 Collaboration,
 Phys. Rev. Lett. \textbf{81},  4309 (1998).

\bibitem{belyaev} A. Belyaev {\it et al.}, hep-ph/0107046.

\end{thebibliography}
\end{document}